\documentclass[prl,aps,floatfix,twocolumn,superscriptaddress,showpacs,10pt]{revtex4-1}
\usepackage{graphicx}
\usepackage{amsmath,amssymb}
\usepackage{bm}
\usepackage[font=small,format=plain,justification=centerlast,singlelinecheck=true]{caption}
\usepackage{subcaption}
\usepackage[colorlinks=true,citecolor=blue,linkcolor=black]{hyperref}
\def\equationautorefname~#1\null{Eq.~(#1)\null}
\def\figureautorefname~#1\null{Fig.~(#1)\null}

\def\<{\left<}
\def\>{\right>}  
\newcommand{\ket}[1]{|{#1}\rangle}
\newcommand{\bra}[1]{\langle{#1}|}
\newcommand{\braket}[1]{\langle #1 \rangle}
\def\elem<#1|#2|#3>{\left<#1\right|#2\left|#3\right>}
\def\({\left(}
\def\){\right)}

\begin{document}
\title{Quantum walks in synthetic gauge fields with 3D integrated photonics}
\author{Octavi Boada}
\affiliation{Physics of Information and Quantum Technologies Group, Instituto de Telecomunica\c{c}\~oes, Lisbon, Portugal}
\author{Leonardo Novo}
\affiliation{Physics of Information and Quantum Technologies Group, Instituto de Telecomunica\c{c}\~oes, Lisbon, Portugal}
\affiliation{Instituto Superior T\'{e}cnico, Universidade de Lisboa, Portugal}
\author{Fabio Sciarrino}
\affiliation{Dipartimento di Fisica, Sapienza Universita di Roma,
Roma, Italy }
\author{Yasser Omar}
\affiliation{Physics of Information and Quantum Technologies Group, Instituto de Telecomunica\c{c}\~oes, Lisbon, Portugal}
\affiliation{Instituto Superior T\'{e}cnico, Universidade de Lisboa, Portugal}
\bigskip

\begin{abstract}
There is great interest in designing photonic devices capable of disorder-resistant transport and information processing. In this work we propose to exploit 3D integrated photonic circuits in order to realize 2D discrete-time quantum walks in a background synthetic gauge field. The gauge fields are generated by introducing the appropriate phase shifts between waveguides. Polarization-independent phase shifts lead to an Abelian or magnetic field, a case we describe in detail. We find that, in the disordered case, the magnetic field enhances transport due to the presence of topologically protected chiral edge states which do not localize. Polarization-dependent phase shifts lead to effective non-Abelian gauge fields, which could be adopted to realize Rashba-like quantum walks with spin-orbit coupling. Our work introduces a flexible platform for the experimental study of multi-particle quantum walks in the presence of synthetic gauge fields, which paves the way towards topologically robust transport of many-body states of photons. 
 \end{abstract}
\date{\today}

\maketitle
\vspace{-50px}
%Introduction
%============
%Electrons in a magnetic field, fundamental interest, applications: topo insulators & topo quant comp
\textit{Introduction}. A long-standing aim in condensed matter physics is to understand the behavior electrons in two dimensional systems in the presence of a magnetic field \cite{klitzing1980new}. The reasons for this are both of fundamental and of applied nature. When the system is well described by weakly interacting quasielectrons, it is known that topologically-protected edge states akin to those of topological insulators \cite{hasan2010colloquium} are present. Strongly interacting electrons in a magnetic field arrange themselves in non-standard states of matter \cite{laughlin1983anomalous} which cannot be described by a local order parameter and present topological order \cite{wen1990ground}. The excitations of this state of matter may present non-Abelian statistics, which could be used for topologically-protected quantum computation \cite{nayak2008non}. 

%Simulation of artificial gauge fields using different platforms, emphasys cold atoms
The promise of ground-breaking applications together with the richness of the underlying physics of two-dimensional ($2D$) quantum particles in a magnetic field has made these systems a favorite subject of quantum simulator proposals \cite{buluta2009quantum}. In these quantum simulators---physical systems unnaturally made to behave according to a specific \textit{model}---the magnetic field is artificial, i.e.\ synthetic. Instead of using charged particles in an actual magnetic field, in a quantum simulator one typically uses neutral particles upon which the effects of a fictitious magnetic field are imposed. For neutral cold atom approaches, methods used to generate a synthetic magnetic field include rapid rotation \cite{madison2000vortex,abo2001observation}, Raman-laser-induced Berry phases \cite{dalibard2011colloquium}, laser-stimulated tunneling in optical lattices \cite{jaksch2003creation,gerbier2010gauge,celi2014synthetic,lin2009synthetic,miyake2013realizing,
aidelsburger2013realization}, or lattice shaking \cite{struck2011quantum}. 

%Quantum walks approach
An alternative approach to quantum simulation is to directly implement the time-evolution of the system, as opposed to engineering the underlying Hamiltonian. Quantum walks (QW) \cite{kempe2003quantum} are a prominent example of this idea and have been realized in a variety of platforms, including neutral trapped atoms \cite{karski2009quantum}, trapped ions \cite{schmitz2009quantum,zahringer2010realization}, and nuclear magnetic resonance in continuous \cite{du2003experimental} and discrete-time \cite{ryan2005experimental}. A promising platform are photonic quantum simulators \cite{aspuru2012photonic}, which have been used to simulate QWs in the bulk \cite{zhang2007demonstration,broome2010discrete} and in waveguide lattices \cite{perets2008realization}, as well as photon time-bin encoded  QWs \cite{schreiber2010photons}. Furthermore, two-particle QWs \cite{omar2006quantum} have been realised in integrated photonic circuits using quasi-planar geometries  \cite{peruzzo2010quantum,owens2011two,sansoni2012two}, non-planar circuits in a ``criss-cross" configuration \cite{poulios2014quantum,crespi2014fermionic}, and Anderson localization has been reported in the disordered case \cite{crespi2013anderson}.

%1D QW with integrated photonics
Discrete-time QWs (DTQWs) in $1D$ may be implemented with a planar integrated photonic circuit (IPC) forming an array of beam-splitters \cite{sansoni2012two}. Each beam-splitter performs the coin and step operator at the same time, shifting the photon left and right in quantum superposition. Successive beam-splitters create further superpositions, leading to the genuinely quantum interference phenomena which are characteristic of QWs. In this implementation, time is encoded in the direction of propagation of the photon in the IPC.

\begin{figure}
\begin{subfigure}[b]{1\columnwidth}
\includegraphics[width=0.6\columnwidth]{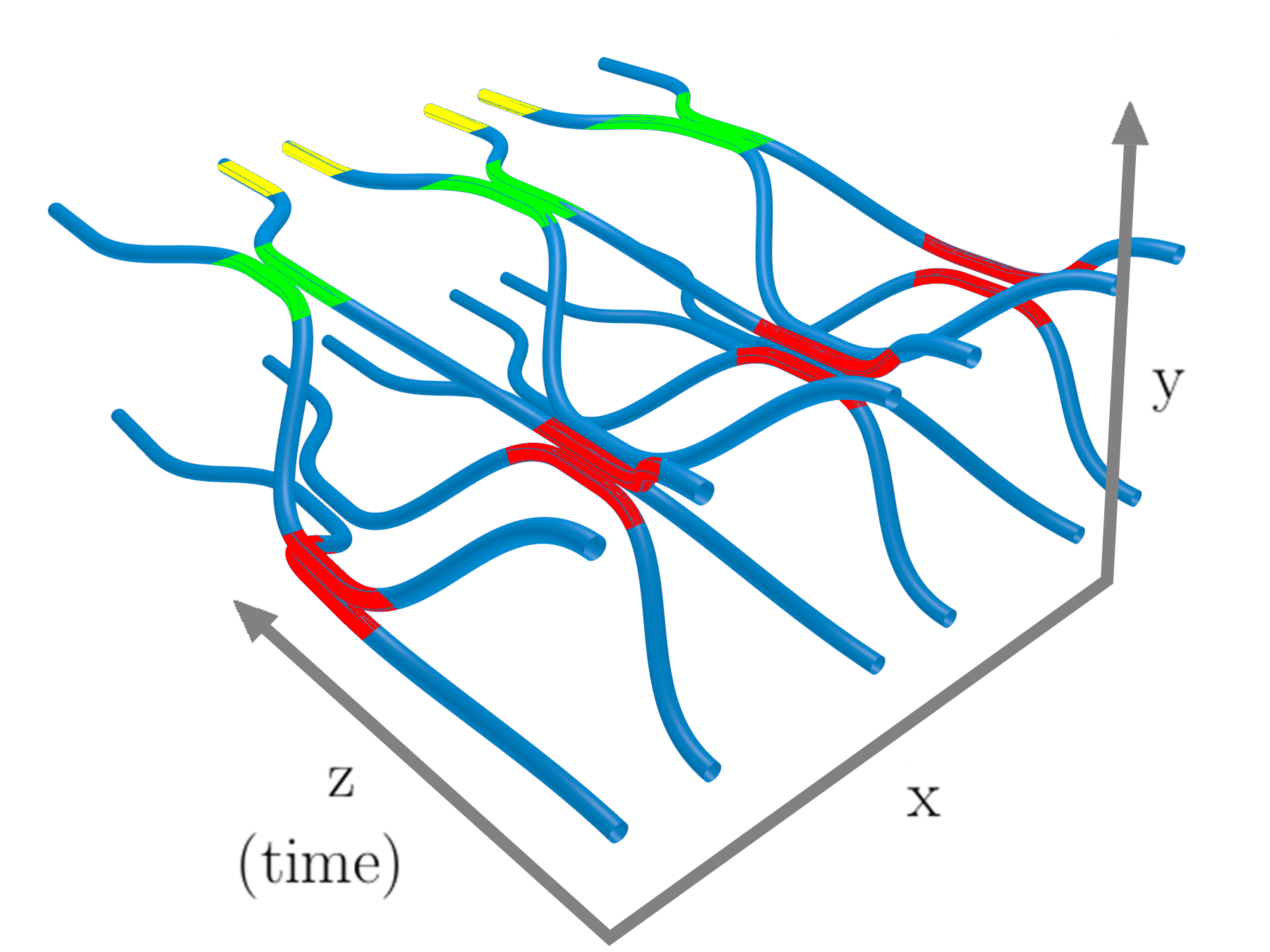}
                  \vspace{-3px} \caption{}\label{circuit}
\end{subfigure} 
\begin{tabular}{c | c} \begin{subfigure}[b]{0.3\columnwidth}
\includegraphics[width=\columnwidth,trim= 0cm 0cm 0cm 0cm ]{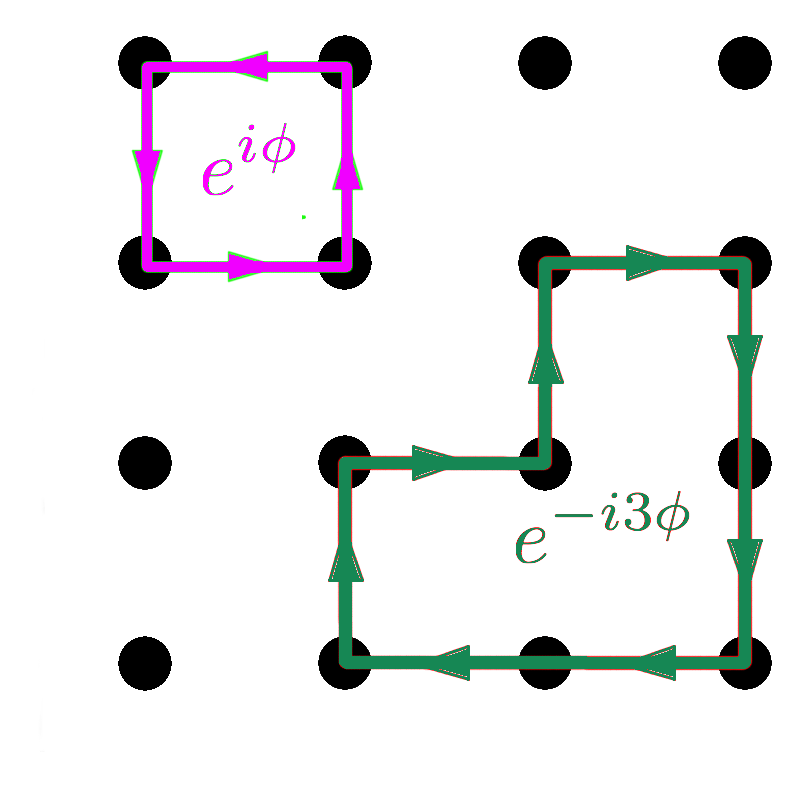}\vspace{-5px} 
               \caption{}\label{aharanovphase}
                \end{subfigure}~~~~~~~&~~~~~~ 
                \begin{subfigure}[b]{0.35\columnwidth}
\includegraphics[width=0.9\columnwidth]{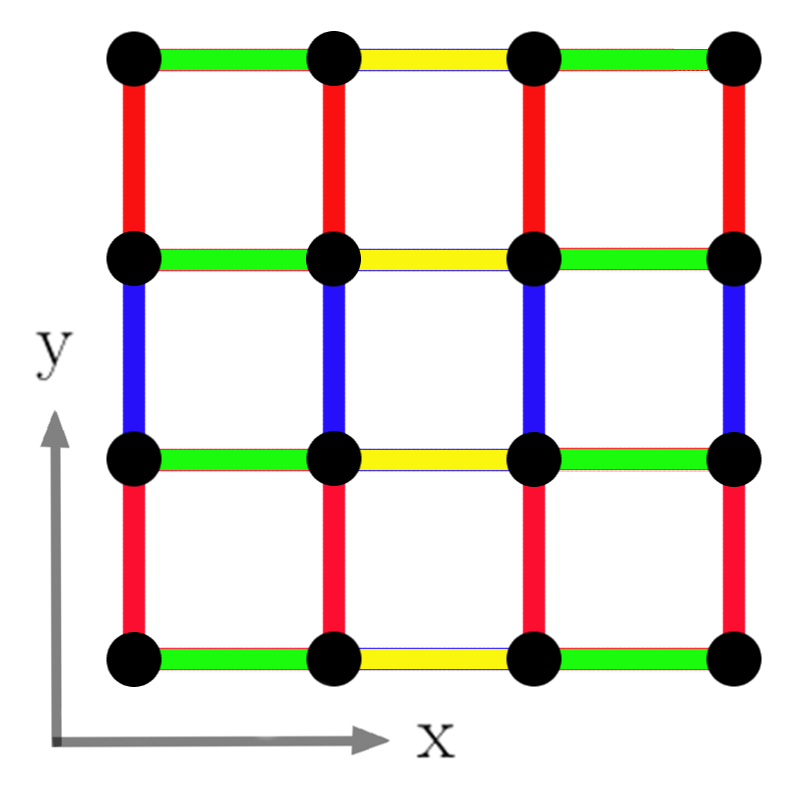}
                \caption{}\label{trotterlinks}
                \end{subfigure}
                \end{tabular}
                \vspace{-6px}
                \caption{(a): Diagram of the proposed $3D$ photonic circuit that realizes the DTQW on a $2D$ lattice. The $z$ axis represents the direction of time, while the $xy$ plane represents the two spatial dimensions where the QW takes place. Each full time-step of the quantum walk is implemented by a sequence of four sub-steps, depicted here and in (c) in four different colors. Each substep implements a set of mutually commuting gates between waveguides which can be applied simultaneously. The couplings in the $y$ direction introduce a $x$-dependent phase-shift between waveguides which mimics the effect of a magnetic field. (b): The accumulated phase acquired by the quantum walker going around two examples of closed trajectories with opposite chirality. This phase depends only on the chirality and the number of elementary cells inside the loop, as in the Aharanov-Bohm effect experienced by a charged particle in a constant magnetic field. (c): The four different kinds of links involved in the quantum walk. Each group of links is depicted in a different color.  The DTQW is realized by creating superpositions between sites in each group in a sequential way.\vspace{-20px}}\label{latdiags}
\end{figure}

A promising development in IPC technology is the capability to print the waveguides in a truly $3D$ configuration. In particular, it is possible to implement quantum walks on a $2D$ lattice using a $3D$ network of beam- splitters. In such a network, each waveguide corresponds to one lattice site, and there are vertical and horizontal beam-splitters, which shift the photon wave function in an up-down and left-right superposition, respectively (see \ref{circuit}). Similarly to the implementation of 1D quantum walks, the time is encoded in the spatial direction of propagation of the photon.

In this work we propose $3D$ integrated circuits to realize $2D$ QWs in a synthetic gauge field. This is accomplished by introducing controlled phase-shifts between waveguides at the beam-splitters. The phase-shifts are chosen in such a way that the photons gain global phases when going around a closed loop, leading to the Aharonov-Bohm effect \cite{aharonov1959significance} (see \autoref{aharanovphase}). Polarization-independent phase shifts lead to an Abelian or magnetic field, while polarization-dependent phase shifts lead to a non-Abelian gauge field \cite{kogut1979introduction}. Our scheme may be readily generalized to QWs of two or more photons \cite{omar2006quantum}, allowing for the implementation of QW exhibiting topological features \cite{kitagawa2010exploring,kitagawa2012observation} in the multi-photon case in an IPC. Furthermore, the spatial dependence of the effective gauge field is highly tuneable, thus allowing for synthetic gauge fields in exotic configurations, such as magnetic monopoles, with no added difficulties.  There is great interest in engineering photonic technologies with topologically protected properties \cite{photoTIreview}. Although several examples of photonic systems with topologically protected edge states have been proposed \cite{haldane2008possible,fang2012realizing,schmidt2015optomechanical} and realized \cite{hafezi2011robust,mittal2014topologically} with laser light, such as the quantum Hall effect and the Floquet topological insulator \cite{floquetSzameit}, our proposal is to realize quantum walks in effective gauge fields in the few-walker regime, using single-photons. \\
%Hamiltonian generating quantum walk, abelian and non-abelian
\textit{ 2D QW in a synthetic magnetic field with an IPC. } The evolution of a charged bosonic particle in a $2D$ lattice with a perpendicular magnetic field is described by the Hamiltonian \vspace{-3px}
\begin{equation}
H=J\sum_{m,n} (e^{i\phi m} a^{\dagger}_{m,n+1} a_{m,n} +a^{\dagger}_{m+1,n} a_{m,n}+H.c.)\, .\label{abelianHam}
\vspace{-5px}
\end{equation}
The operators $a^{\dagger}_{m,n}$ and $a_{m,n}$ create and destroy one particle at site $(m,n)$ of the lattice, respectively, and obey bosonic commutation relations. The constant $J$ is an arbitrary energy scale and $\phi$ is the magnetic flux per plaquette. The key feature of this Hamiltonian is that hopping in one of the directions of the lattice entails the acquisition of a position-dependent phase. The specific spatial profile of these phases is such that the global phase acquired by a particle going around a closed path on the lattice is position-\textit{ independent} and equal to $e^{i\phi N}$, where $N$ is the number of elementary cells inside the path (see \autoref{aharanovphase}). The particular choice of phases is arbitrary---in \autoref{abelianHam} we chose the so-called Landau gauge for convenience---as long as the accumulated phase along closed paths leads to the correct global phase. The idea of introducing position-dependent phases has previously been used to realize chiral QWs on graphs \cite{zimboras2013quantum,lu2014chiral}, as well as a QW in an effective electric field \cite{cedzich2013propagation,genske2013electric}.

Here, we use an approach involving coinless discrete-time quantum walks on a 2D lattice where each step implements a position-dependent phase, in analogy with the dynamics given by Hamiltonian from Eq.~\eqref{abelianHam}. We explain how to implement this quantum walk in a 3D IPC and present numerical evidence that its dynamics shows similar features to the one described by \eqref{abelianHam}, namely the presence of topologically protected edge states. Although in other photonic implementations of discrete-time  quantum walks the polarization of the photon is used as the coin \cite{broome2010discrete,schreiber2010photons}, here we assume the IPC to be polarization independent so that we can use entanglement in the polarization to simulate bosonic and fermionic statistics\cite{omar2006quantum}, as previously done in 1D quantum walks in Refs.~\cite{crespi2013anderson}, and so we have a coinless quantum walk. For a general definition and discussion of spreading properties of coinless quantum walks on lattices see Ref.~\cite{Portugal}.

To implement the quantum walk on the 2D lattice using a 3D IPC, a lattice site in a position (x,y) will correspond to a waveguide engraved in the IPC, also labelled by the position (x,y), which is extended in the z-direction, corresponding to the time dimension of the DTQW. The hops of the quantum walk correspond to sequences of beam splitters, which can be implemented by bringing adjacent waveguides close together in such a way that they are evanescently coupled. Each lattice site has at most 4 neighbors but, since we can only couple two waveguides at a time, it is not possible for the photon in a certain waveguide to hop to all its nearest neighbors in one step. This way, we divide the DTQW into 4 substeps, as depicted in Fig.~\ref{trotterlinks}, where the links in green correspond to hopping terms that will be implemented by a unitary matrix $U_1$, the ones in yellow by $U_2$, the ones in red by $U_3$ and finally the ones in blue by $U_4$. This way, one full step of the quantum walk will be given by \vspace{-10px}
\begin{equation}
U_{\text{step}}=U_4 U_3 U_2 U_1.
\vspace{-7px}
\end{equation}
This sequence of unitaries can be applied many times along the z-direction of the IPC to implement subsequent steps of the quantum walk. A schematic representation of part of the 3D photonic circuit implementing the red, green and yellow links is shown in Fig.~\ref{circuit}.

In order to mimic the effect of a magnetic field, we need to construct $U_{\text{step}}$ in such a way that if the walker makes a closed path around N elementary cells, it acquires a position-independent phase $e^{i \phi N}$. To show the form of the unitaries $U_i$ which satisfy this requirement, we define the states of the Hilbert space $\ket{x}\ket{y}$, with $x\in \{1,M\}$ and $y\in \{1,M\}$, and say that the photon is in state $\ket{x}\ket{y}$ if it is in the waveguide labelled by the coordinates (x,y). In this basis, we define the hopping operators in the $x$ and $y$ directions as
\begin{align}
V_x=&\frac{1}{\sqrt{2}}\left( \ket{x}\bra{x}+\ket{x+1}\bra{x+1}\right)\\
&+\frac{i}{\sqrt{2}}\left( \ket{x}\bra{x+1}+\ket{x+1}\bra{x}\right)\nonumber\\
V_y(\phi)=&\frac{1}{\sqrt{2}}\left( \ket{y}\bra{y}+\ket{y+1}\bra{y+1}\right)\\
&+\frac{i}{\sqrt{2}}\left(e^{-i \phi} \ket{y}\bra{y+1}+e^{i \phi}\ket{y+1}\bra{y}\right),\nonumber
\end{align}
which corresponds to an unbiased beam-splitter matrix and to a phase-shifted beam-splitter, respectively. The operators $U_1$, $U_2$, $U_3$ and $U_4$ are then defined as 
\begin{align}
U_1=&\sum_{x=0}^{M/2-1}V_{2x+1}\otimes \mathcal{I}_y,~~~U_2=\sum_{x=1}^{M/2-1}V_{2x}\otimes \mathcal{I}_y\nonumber \\
U_3=&\sum_{x=1}^{M}\sum_{y=0}^{M/2-1}\ket{x}\bra{x}\otimes V_{2y+1}(x \phi)\\
U_4=&\sum_{x=1}^{M}\sum_{y=1}^{M/2-1}\ket{x}\bra{x}\otimes V_{2y}(x \phi).\nonumber
\end{align}
$U_3$ and $U_4$ cause the hopping of the photon in the y-direction and apply a phase which is proportional to the coordinate $x$. Previous experiments have shown full phase-shift controllability between two waveguides, by deforming one of the waveguides and thus creating a difference in the optical path length \cite{crespi2013anderson}. Hence, the experimental implementation of $U_{\text{step}}$, although challenging, is within reach of current technology.
%\begin{figure}[htb]
%\begin{centering}
%\includegraphics[width=.8\columnwidth]{./figures/spectrum.png} \tabularnewline
%\end{centering}
%\caption{Hofstadter's butterfly in a finite $2D$ lattice with open boundary conditions: spectrum of the effective Hamiltonian $H_{eff}=i\log U_{IPC}$ which generates the Abelian QW proposed here, as a function of the magnetic flux per plaquette, $\phi$. The unitary operator \textcolor{red}{$U_{IPC}$} which implements the quantum walk is the first-order Suzuki-Trotter expansion of the time evolution generated by $H$.}\vspace{-15px}\label{optical_trotterized}
%\end{figure}
\begin{figure}[h]
        \centering
        \begin{subfigure}[b]{0.49
       \columnwidth}
\includegraphics[width=\columnwidth,trim = 0.1cm 0pt 1.1cm 0pt, clip=true]{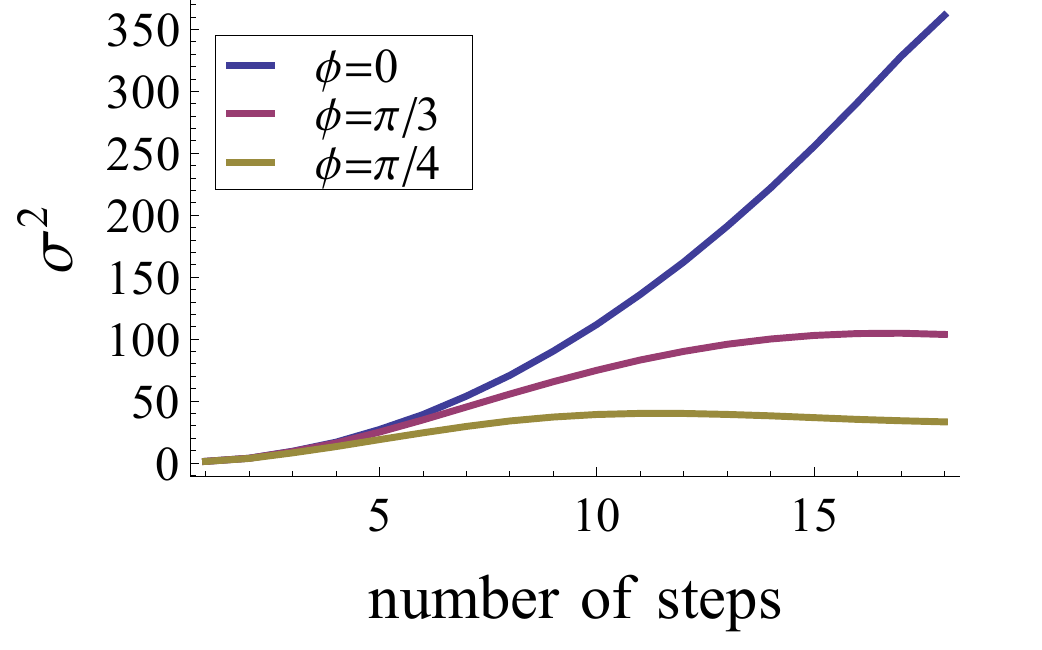}\vspace{-3pt}
                \caption{}\label{var}
                \end{subfigure}
                \begin{subfigure}[b]{0.49\columnwidth}               
\includegraphics[width=\columnwidth,trim = 0.4cm 0pt 1.5cm 0pt, clip=true]{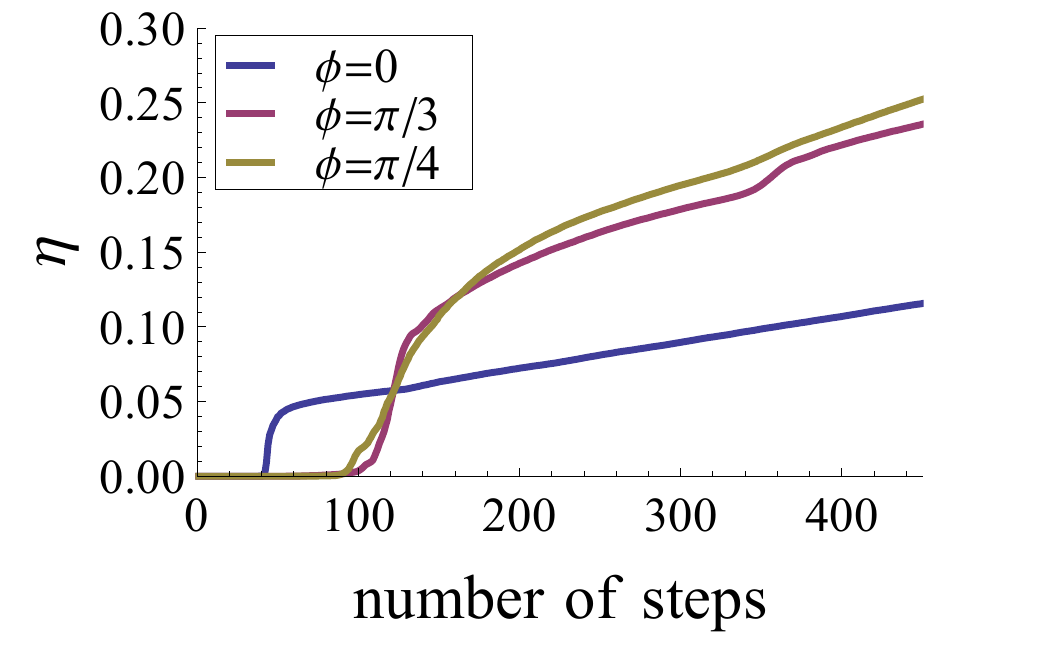}\vspace{-3pt}       
                \caption{}\label{trans}
                \end{subfigure}
\caption{In an ordered lattice, a magnetic field hinders the spreading of the wave function and quantum transport; in a disordered lattice, a magnetic field enhances transport due to the presence of non-localized edge states. The variance of the single-photon wave function vs. number of time-steps, for different values of the magnetic flux $\phi$ is displayed in (a). The inital state is localized at the center of the lattice. For non-zero $\phi$ the QW spreads at a slower rate. In (b), transport efficiency $\eta$ as function of time in the presence of Gaussian disorder with strength $\delta=0.1$ (defined in the Appendix). The quantity $\eta$ measures the accumulated norm at the opposite corner, or the efficiency of quantum transport across the lattice. In the presence of disorder, the applied magnetic field qualitatively enhances transport.\vspace{-10px}}\vspace{-10px}
\end{figure}
\\	
\textit{Signatures of the magnetic field for single photons. }
To confirm qualitatively the correspondence between the proposed IPC and the time evolution generated by \autoref{abelianHam}, we have computed the spectrum of the effective Hamiltonian $H_{eff}=i\log U_{\text{step}}$, where $U_{\text{step}}$ is the unitary operator implemented by the proposed optical circuit. In the Appendix we plot the spectrum of $H_{eff}$ as a function of $\phi$. We obtain a figure very similar to Hofstadter's butterfly \cite{hofstadter1976energy}.

Furthermore, we investigate the effect of the synthetic magnetic field on the spreading and transport properties of the QW at the single-photon level, with and without disorder. Controlled disorder may be implemented via small, random differences in waveguide lengths at the evanescent couplings \cite{crespi2013anderson}, which lead to fluctuations in each waveguide's optical path (see Appendix). These fluctuations are static, in the sense that they are not time-dependent ($z$-dependent in \autoref{circuit}).

To determine how quickly the QW spreads without disorder, in \autoref{var} we plot the variance of the single-particle probability distribution $\sigma^2$ as a function of the number of steps, for different values of $\phi$, the magnetic flux per plaquette. The initial wave function is localized at the center of the lattice. Although we plot here the result for three values of $\phi$, we have observed that the variance is always smaller for $\phi \neq 0$ than for $\phi=0$. Hence, without disorder the magnetic field is detrimental to the expansion of an initially localized photon wave function. 
We also study the QW evolution by computing the transport efficiency between two far-apart waveguides in the presence of disorder. We choose one corner of the lattice as an initial site/waveguide, and the site at the opposite corner as the target. We introduce absorption at the target waveguide, corresponding to position $(M,M)$, at each step of the QW by replacing the operator $U_{\text{step}}$ by $U_{\text{step}}(\mathcal{I}-\ket{M}\ket{M}\bra{M}\bra{M})$. Our measure of transport efficiency, $\eta$, is the accumulated probability of finding the photon at the target, $\eta=\sum_t \vert \braket{\text{target}\vert\psi(t)} \vert^2$, an approach similar to that used in \cite{PhysRevA.91.022324}. Let us stress that $\eta$ could be measured in an experiment, by coupling the target waveguide to a long chain of waveguides as proposed in \cite{crespi2014fermionic}, or to a detector at every time step.

With disorder, low transport efficiency is expected, due to Anderson localization. Interestingly, we find that while in the ordered case a magnetic field slows down the expansion of the QW, in the case of disorder it does the opposite, thus enhancing quantum transport (see \autoref{trans}). This is attributable to the presence of chiral edge states (see Fig. (3)  in Appendix ) which are topologically protected against localization.

\begin{figure}[t]
        \centering
\includegraphics[width=0.7\columnwidth, trim= 0pt 0pt 0pt 7pt, clip=true]{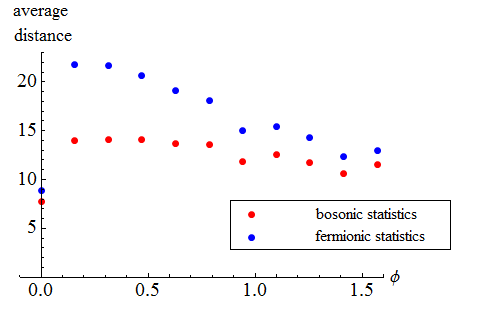}
 \vspace{-0.2cm}                 
%                \quad\begin{subfigure}[b]{\columnwidth}
%               
%\includegraphics[width=0.3\columnwidth]{././figures/plot_probedge.pdf}
% \vspace{-0.6cm}                
%                \caption{}\label{trans}
%                \end{subfigure}
\caption{Average distance between photons for the two photon quantum walk on the IPC, after 20 steps, in a lattice of size 30x30, for different values of the magnetic flux $\phi$. The initial state is localized at positions (1,1) and (1,2) of the lattice and the photons' polarization states are entangled in a symmetric (antisymmetric) way so that the exchange statistics of the wave-function is bosonic (fermionic). We observe, as expected, that for bosonic statistics the photons tend to remain closer than in the fermionic case. Also, the presence of the magnetic field increases the distance between the photons.\vspace{-20px}}
\label{2photonplots}
\end{figure}

\textit{2-photon QW with a magnetic field.}
The single-particle probabilities obtained by using one photon as the input state of the IPC can be reproduced by using a classical laser light source. However, if two or more indistinguishable photons are used as input, the probability distribution measured at the output of the circuit has no classical analogue and for many photons is, in general, hard to calculate \cite{aaronson,crespi2013integrated}. Also, by choosing appropriate entangled states of two-photons, the statistics of bosons and fermions can be mimicked \cite{omar2006quantum} and bunching and anti-bunching phenomena have been observed in 1-D QW's \cite{crespi2013anderson}. Here, we compute some observables for the DTQW's of two entangled photons in a synthetic magnetic field. The average distance between photons is plotted in Fig.~\ref{2photonplots} for two entangled photons starting at the corner of the lattice. It is clear the effect of the particle statistics in this quantity since bosons remain closer than fermions. Also, the presence of the magnetic field increases the average distance between particles. The presence of two-particle edge states can be seen from the probability that both photons are at the edge of the lattice shown in the Appendix.
\\ \textit{ Non-Abelian 2D QW}. Our proposed scheme to realize a magnetic QW with IPC may be generalized to a non-Abelian magnetic QW, provided the relative phases between adjacent waveguides are made polarization-dependent in a controlled way. When polarization is taken into account, a general term coupling two adjacent lattice sites $i$ and $j$ can be written in the form $\sum_{\xi\tau} a^{\dagger}_{i\xi}U_{ij}^{\xi\tau}a_{j\tau}+H.c.\, ,$
where $\xi,\tau$ run over photon polarizations, and now $a^{\dagger}_{i\xi}$ creates a photon in site $i$ with polarization $\xi$. 

Thus, to realize a QW in a non-Abelian synthetic gauge field, the beam-splitter matrices must be polarization-dependent, which are now described by $4\times 4$ matrices instead of $2\times 2$. In general, the beam-splitter matrices corresponding to different links of the lattice will not commute with each other, and will lead to non-trivial non-Abelian fluxes when the photons go around a closed loop (see \autoref{aharanovphase}). This is tantamount to a modified Aharonov-Bohm effect, where the photon wave function is multiplied by the Wilson loop \cite{peskin1995introduction} instead of a phase. 

Remarkably, interesting non-Abelian QWs may be implemented using relatively simple IPCs. In particular, a QW with Rashba spin-orbit coupling \cite{rashba1960properties} may be realized with the choice $U_x=\exp (i\alpha \sigma_y)$ and $U_y=\exp (-i\alpha \sigma_x)$, where $\sigma_x$ and $\sigma_y$ are Pauli matrices. In the $3D$ IPC architecture, this means adjacent waveguides in the $x$ direction are coupled with $U_x$ and those in the $y$ direction with $U_y$. Note that this choice does not require position dependent delays between waveguides as the Abelian magnetic field case does. In this scenario, since the circuit is now polarization-dependent, it would not be possible to simulate different particle-statistics by entangling photons in polarization.

\textit{ Conclusion}. We have introduced a scheme that allows implementing quantum walks in synthetic gauge fields using integrated photonic circuits. This scheme requires a strong experimental and technological effort: we need the capability to engineer 3D structures with a significant number of steps. In the last year several improvements have been achieved: 8 mode fast Fourier transform with 3D structure \cite{crespi2015}, reconfigurable phase \cite{fulvio} and operation at telecom wavelength which ensures lower losses and hence the possibility to realize longer chips \cite{fulvio}. Our proposal is well suited for the study of topological insulators at the single and few photon levels and it is highly flexible, allowing for the simulation of both Abelian and non-Abelian gauge fields. We have studied the single-photon quantum walk in a constant Abelian or magnetic field and computed experimentally-accessible observables, demonstrating topological properties, namely the presence of  edge states enhancing transport across disordered lattices. We have also computed observables for two-particle quantum walks that demonstrate the role of entanglement and magnetic field in the behaviour of the walk. Overall, we have shown that the development of 3D integrated photonics can lead to the experimental study of interesting 2D quantum physics  with topological features in the few-body regime.

%%============
\begin{acknowledgments}
\textit{ Acknowledgments}. O.B.\ , L.N.\ and Y.O.\ acknowledge support from Funda\c{c}\~{a}o para a Ci\^{e}ncia e a Tecnologia (Portugal), namely through programmes PTDC/POPH/POCH and projects UID/EEA/50008/2013, IT/QuSim, ProQuNet, partially funded by EU FEDER, and from the EU FP7 project PAPETS (GA 323901). Furthermore LN acknowledges the support from the DP-PMI and FCT (Portugal) through scholarship SFRH/BD/52241/2013. F.S.\ acknowledges support from the ERC Starting Grant 3D-QUEST (3D Quantum Integrated Optical Simulation; Grant Agreement No. 307783). 
\end{acknowledgments}

%%%%%%%%%%%%%%%%%%%%%%%%%%%%%%%%%%%%%%%%%%%%%%%%%%%%%%%%%%%%%%%%%%%%%
%
%\bibliographystyle{plain}
\bibliography{bib_gauge.bib}
\onecolumngrid
\appendix
\section*{Appendix}
\label{appendix}
\section{Spectrum of the unitary implemented by the Integrated Photonic Circuit}
\begin{figure}[htb]
\begin{centering}
\includegraphics[width=.45\columnwidth]{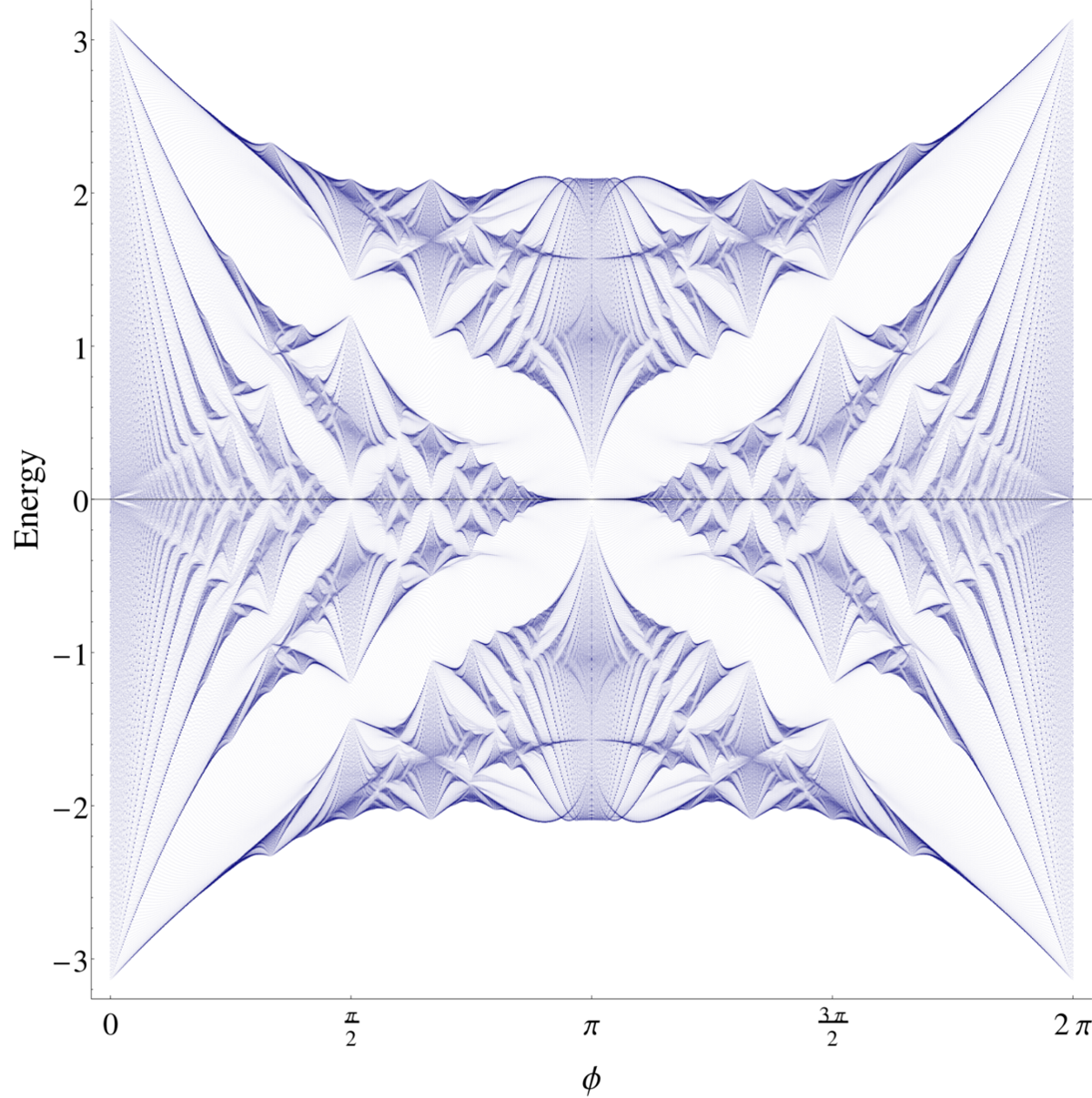} \tabularnewline
\end{centering}
\caption{Spectrum of the effective Hamiltonian $H_{eff}=i\log U_{\text{step}}$ which generates the Abelian QW proposed here, as a function of the magnetic flux per plaquette, $\phi$.}\vspace{-15px}\label{optical_trotterized}
\end{figure}

We have constructed a unitary matrix which implements one time step of the discrete-time quantum walk (DTQW) with a synthetic magnetic field, defined by Eq.~(2) of the main text. This matrix, denoted as $U_{\text{step}}$, can be decomposed in a product of beam-splitter and phase shifter matrices, which can be implemented in an Integrated Photonic Circuit (IPC). In \ref{optical_trotterized} we plot the spectrum of the effective Hamiltonian $H_{eff}=i\log U_{\text{step}}$ as a function of $\phi$. As expected, we obtain a figure very similar to Hofstadter's butterfly \cite{hofstadter1976energy}---a complex, self-similar structure which arises in the case of electrons propagating on a $2D$ lattice in a strong magnetic field. For relatively large lattices (30 x 30), the spectrum of $H_{eff}$ presents a fractal nature and a structure of gaps which is very reminiscent of Hofstadter's butterfly.

\section{Time-evolution of single-particle quantum walks in a magnetic field}
\begin{figure}[h!]
        \centering
        \begin{subfigure}[b]{0.32\textwidth}
                \includegraphics[width=\textwidth]{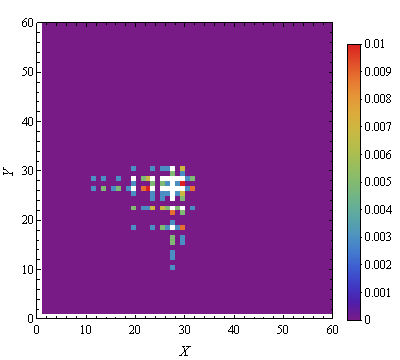}
                \caption{20 steps}
                \end{subfigure}%
        ~ %add desired spacing between images, e. g. ~, \quad, \qquad, \hfill etc.
        \begin{subfigure}[b]{0.32\textwidth}
                \includegraphics[width=\textwidth]{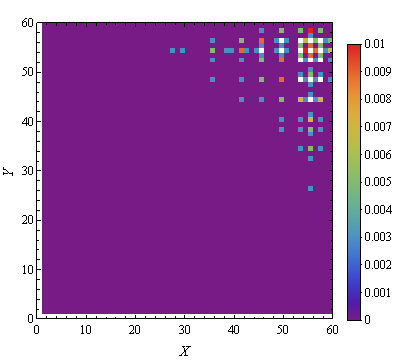}
                \caption{40 steps}
                \end{subfigure}
        ~ %add desired spacing between images, e. g. ~, \quad, \qquad, \hfill etc.
        \begin{subfigure}[b]{0.32\textwidth}
                \includegraphics[width=\textwidth]{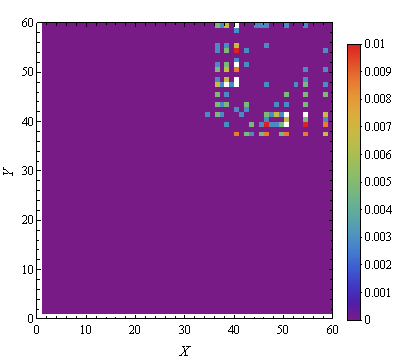}
                \caption{60 steps}
                \end{subfigure}
        \caption{Simulation of the IPC implementation of a DTQW in a $2D$ lattice in zero magnetic field without disorder, starting at position (X,Y)=(1,1). Probability distribution of a discrete-time QW on a 60x60 ordered lattice after 20, 40 and 60 time steps. The quantum walk propagates quickly from one corner of the lattice to the opposite corner. The white color corresponds to probabilities above $0.01$, which can go up to $0.11$ in (a), 0.056 in (b) and 0.016 in (c). Total probability for each plot: (a) 1  , (b) 1, and (c) 0.77. }\label{ord_QW}
\end{figure}

\begin{figure}[h!]
        \centering
        \begin{subfigure}[b]{0.32\textwidth}
                \includegraphics[width=\textwidth]{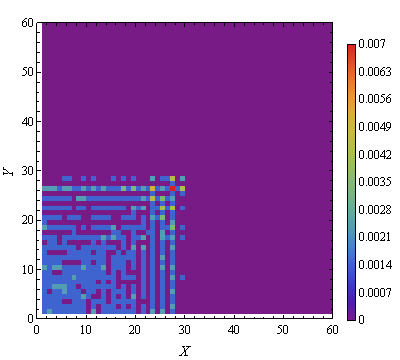}
                \caption{20 steps}
                \end{subfigure}%
        ~ %add desired spacing between images, e. g. ~, \quad, \qquad, \hfill etc.
        \begin{subfigure}[b]{0.32\textwidth}
                \includegraphics[width=\textwidth]{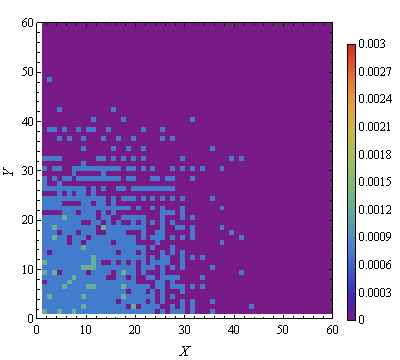}
                \caption{40 steps}
                \end{subfigure}
        ~ %add desired spacing between images, e. g. ~, \quad, \qquad, \hfill etc.
        \begin{subfigure}[b]{0.32\textwidth}
                \includegraphics[width=\textwidth]{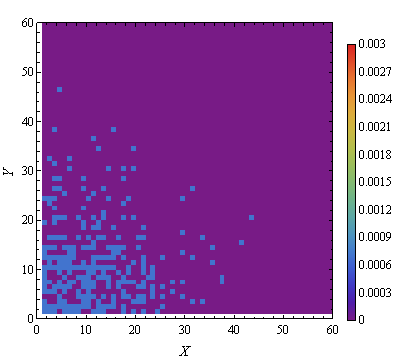}
                \caption{60 steps}
                \end{subfigure}
        \caption{Simulation of the IPC implementation of a DTQW in a $2D$ lattice in zero magnetic field with disorder, starting at position (X,Y)=(1,1). Probability distribution of a discrete-time QW on a 60x60 disordered lattice after 20, 40 and 60 steps. The strength of disorder is $\sigma=0.2$ and the magnetic field is set to zero. The presence of disorder hinders the propagation on the lattice and the photon remains localized close to its starting position, resulting in a low transport efficiency to the opposite corner. The probability amplitude is averaged over 20 disorder realizations. Note that the scales of the plots are different, for better visualization. Total probability for each plot: (a) 1, (b) 1, and (c) 0.997.}\label{dis_QW}
        \vspace{-10px}
\end{figure}

\begin{figure}[h!]
        \centering
        \begin{subfigure}[b]{0.32\textwidth}
                \includegraphics[width=\textwidth]{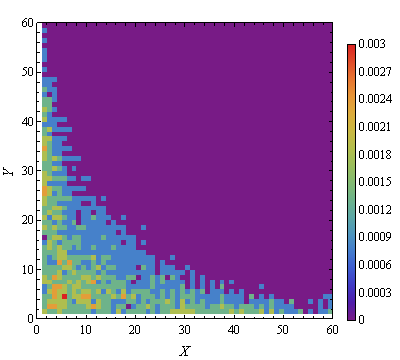}
                \caption{50 steps}
                \end{subfigure}%
        ~ %add desired spacing between images, e. g. ~, \quad, \qquad, \hfill etc.
        \begin{subfigure}[b]{0.32\textwidth}
                \includegraphics[width=\textwidth]{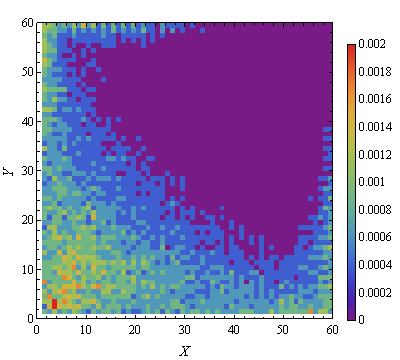}
                \caption{100 steps}
                \end{subfigure}
        ~ %add desired spacing between images, e. g. ~, \quad, \qquad, \hfill etc.
        \begin{subfigure}[b]{0.32\textwidth}
                \includegraphics[width=\textwidth]{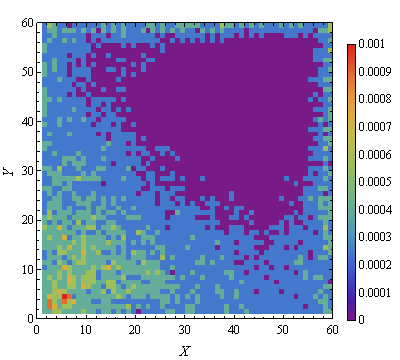}
                \caption{150 steps}
                \end{subfigure}
        \caption{Simulation of the IPC implementation of a DTQW in a $2D$ lattice in non-zero magnetic field with disorder, starting at position (X,Y)=(1,1). Probability distribution of a discrete-time QW on a 60x60 disordered lattice with magnetic field after 50, 100 and 150 steps. The strength of disorder is $\sigma=0.1$ and the magnetic field is $\phi=\pi/5$. The quantum walk propagates mainly along the edges of the lattice. The presence of topologically protected edge states that do not localize allows for the propagation of the photon to the opposite corner of the lattice. Here we investigate a larger number of steps as compared to Figs.~\ref{ord_QW} and \ref{dis_QW}, in order to see the propagation to the opposite corner of the lattice, (X,Y)=(60,60).  The probability amplitude is averaged over 20 disorder realizations. Note that the scales of the plots are different, for better visualization. Total probability for each plot: (a) 1, (b) 1, and (c) 0.984. }\label{dismag_QW}
\end{figure}

We study the effect of a synthetic Abelian gauge field on the transport properties of a discrete-time $2D$ QW in the presence of disorder. All simulations have been done with the beam-splitter matrices given in Eq. (4) and Eq. (5) of the main paper. The initial wave function is localized at the lower left corner of the lattice. In \autoref{ord_QW} we plot the evolution of the QW without magnetic field and without disorder. There is efficient transport from one corner of the lattice to the other. In \autoref{dis_QW}, localization close to the initial position is observed for zero magnetic field and disorder; transport from one corner of the lattice to the other is highly inefficient in this case. In \autoref{dismag_QW} we plot the evolution of the QW for non-zero magnetic field, $\phi=\frac{\pi}{5}$, and in the presence of disorder. Transport from one corner of the lattice to the opposite one is clearly accomplished by edge states, which do not penetrate significantly into the bulk of the lattice. 
%%%

At each step of the QW, the probability at the target waveguide is subtracted from the wave function, as we introduce absorption by replacing the operator $U$ with $U \exp(-a^{\dagger}_{target}a_{target})$. Thus, the deviation from unity of the total probability is equal to the transport efficiency, $\eta=\sum_t \vert \braket{\text{target}\vert\psi(t)} \vert^2$.

The introduction of static disorder in the DTQW in a $2D$ lattice is done by multiplying each beam-splitter matrix $V_x$ or $V_y(\phi)$, defined in Eqs.~(3) and (4) of the main text, by a matrix with random phases in the diagonal, 
\begin{equation}\label{hadtrick_phase_app}
V^{\text{dis.}}_{x,y}=\begin{pmatrix}
e^{i \epsilon_i}&0\\
0 &e^{i \epsilon_j}\end{pmatrix}
V_{x,y}.
\end{equation}
The quantities $\epsilon_i$ are sampled from a normal distribution with standard deviation $\delta$, which we will refer to as the disorder strength. This will lead to a unitary $U'_{\text{step}}$ defining the step of the quantum walk with static disorder. Note that time-dependent disorder, i.e. if $U'_{\text{step}}$ depends on the step number, would lead to dephasing of the QW \cite{crespi2013anderson}.  
\section{Evidence of two-photon edge states}
\begin{figure}[h!]
        \centering
\includegraphics[width=0.5\columnwidth]{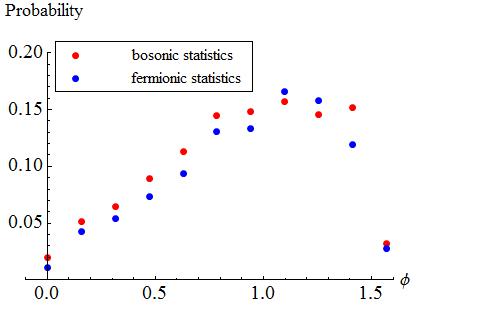}
 \vspace{-0.3cm}                
\caption{Probability that the two photons are at the edge of the lattice for the quantum walk on the IPC, after 20 steps, in a lattice of size 30x30, for different values of the magnetic flux $\phi$. If the photons' polarization states are entangled in a symmetric (antisymmetric) way, their statistics is bosonic (fermionic). The presence of the magnetic field increases significantly the probability that the two photons are at the edge. However, the exchange statistics of the two-photon wavefunction does not affect much this quantity.}\vspace{-10px}
\label{2photonedge}
\end{figure}

A quantity that can easily be calculated from the probability distribution of the position of the photons at the output of the IPC is the probability that the photons leave the circuit in a waveguide which belongs to the edge of the 2D lattice. For the two photon quantum walk, we calculate this probability for a lattice of size 30x30, after 20 steps, and for variable magnetic flux $\phi$ (see Fig.~\ref{2photonedge}). The two photons are inserted in the circuit in the corner of the lattice, position $(X,Y)=(1,1)$, and its nearest neighbour in the x-direction at the position $(2,1)$. The two photons are entangled in polarization in a symmetric (antisymmetric) way in order to simulate bosonic (fermionic) statistics \cite{omar2006quantum,crespi2013anderson}. We see that, for zero magnetic field, it is very unlikely that the photons leave the circuit by the edge of the lattice but with magnetic field this probability increases up to $\approx 15\%$. It is interesting that the particle statistics does not affect much this probability unlike what happens with the average distance between particles shown in the main text.  
%%%%%%%%%%%%%%%%%%%%%%%%%%%%%

\end{document}